\pdfoutput=1
\documentclass{sig-alternate}

\usepackage{moreverb}
\usepackage[square,sort,comma,numbers]{natbib}
\usepackage{amsmath,amssymb, mathtools}

\usepackage{amsthm}

\usepackage{graphicx}
\graphicspath{ {./} }

\def\P{\mathbb{P}}

\newcommand\BibTeX{{\rmfamily B\kern-.05em \textsc{i\kern-.025em b}\kern-.08em
T\kern-.1667em\lower.7ex\hbox{E}\kern-.125emX}}

\makeatletter
\renewcommand{\@ptsize}{0}
\makeatother
\begin{document}

\title{A Class of Temporal Hierarchical Exponential Random Graph Models for Longitudinal Network Data}
%\setcopyright{acmcopyright}
%\CopyrightYear{2017}
%\conferenceinfo{KDD '17,}{August 13 - 17, 2017, Halifax, Nova Scotia, Canada}
%\isbn{978-1-4503-4232-2/16/08}\acmPrice{\$15.00}
%\doi{http://dx.doi.org/10.1145/2939672.2939754}

\numberofauthors{1}
\author{
	\alignauthor
	Ming Cao\\
	\affaddr{University of Texas Health Science Center at Houston}\\
	\email{ming.cao@uth.tmc.edu}
	%\alignauthor Yong Chen\\
	%\affaddr{University of Pennsylvania}\\
	%\email{ychen123@mail.med.upenn.edu}
}

\maketitle

\begin{abstract}
As a representation of relational data over time series, longitudinal networks provide opportunities to study link formation processes. However, networks at scale often exhibits community structure (i.e. clustering), which may confound local structural effects if it is not considered appropriately in statistical analysis. To infer the (possibly) evolving clusters and other network structures (e.g. degree distribution and/or transitivity) within each community, simultaneously, we propose a class of statistical models named Temporal Hierarchical Exponential Random Graph Models (THERGM). Our generative model imposes a Markovian transition matrix for nodes to change their membership, and assumes they join new community in a preferential attachment way. For those remaining in the same cluster, they follow a specific temporal ERG model (TERGM). While a direct MCMC based Bayesian estimation is computational infeasible, we propose a two-stage strategy. At the first stage, a specific dynamic latent space model will be used as the working model for clustering. At the second stage, estimated memberships are taken as given to fit a TERG model in each cluster. We evaluate our methods on simulated data in terms of the mis-clustering rate, as well as the goodness of fit and link prediction accuracy.
\end{abstract}


\footnotetext[2]{}

\section{Introduction}
Relational data over a series of time points are becoming more available as the information technology advances, for example, \cite{salathe2010high} study flu transmission by using sensors to record human contact in a high school. One of the most attractive topics in network research is \textit{community detection} or \textit{clustering} (see \cite{fortunato2010community} for a review). Speaking of scientific literatures, communities could be corresponding to \textit{research fields}. Those \textit{fields} may or may not be pre-specified, take the ``network research community'' as example, it has seen mixed contributions and gaining momentums from scholars in Physics, Computer Science, Sociology, Statistics and other disciplines \citep{newman2011structure,wasserman1994social,jackson2008social}.  Multiple snapshots, or even at the finest grain as a stream of relational events, provide more information than a static network. It provides us opportunities to capture the trajectories of evolving network structures, with applications to trend analysis \citep{leskovec2007patterns} and link prediction \citep{sarkar2014}.\\
Besides the community structure, there could be various types of network effects going on \textit{locally}. For example, the common sense that ``my friends' friends are more likely to be my friends'' and ``people with same interests are more likely to be friends'' are typically referred to as \textit{transitivity} and \textit{homophily}, respectively. With a long history of being studied by Sociologists \citep{granovetter1973strength,mcpherson2001birds}, such kind of network structures can be flexibly specified and therefore estimated/tested by a powerful statistical tool called Exponential Random Graph Models (ERGM) \citep{frank1986markov,wasserman1996logit}. We should have reasonable suspect that transitivity or homophily acts equally in different communities and it would be natural to construct a statistical model that consider local structural effects as well as clustering, which represents a more macro level structure. However, current works for large-scale network structure analysis have mostly been limited to base on the Stochastic Block Models (SBM) \citep{holland1983stochastic,snijders1997block}, which assumes no dependence among ties within the same block (cluster). SBM is adopted for its simplicity in the hope that it approximates well from the perspective of an efficient clustering estimator. Nice statistical properties have been established for computationally effective algorithmic methods such as spectral clustering, under both the static \citep{rohe2011spectral,qin2013regularized,jin2015score} and dynamic \citep{han2014consistent} settings. While efforts are being made to extend this often over-simplified family of models to a boarder class \citep{wan2015class}, ERGM is generally considered having degeneracy problem \citep{handcock2003assessing} and too expensive to compute the MCMC approximated MLE \citep{snijders2002markov}. In fact, significant improvement on model specification and estimation technique has been made \citep{snijders2006new,hunter2006curved,hummel2012improving} for the past decade. More importantly, the longitudinal extensions of ERGM rarely suffer from the model degeneracy problem \citep{hanneke2010temporal}. It can be intuitively understood as that structural effects became much more clear when the changes are explicitly modeled, rather than being very likely blurred in a cumulative snapshot.\\
In this paper, we propose a class of Temporal Hierarchical Exponential Random Graph Models (THERGM) to fully utilize the information contained in longitudinal network data, which may greatly improve the overall goodness of fit and link prediction accuracy. It seems that we pay the price of a non-scalable inference procedure in order to gain the modeling power of ERGM, but fortunately, \cite{leskovec2009size} suggests that roughly 100 nodes is probably a universal "natural size" of a meaningful cluster, including social, information, Internet and biological networks. While this is the right scale that current ERGM implementation \citep{handcock2008statnet} could be routinely carried out, it is still computationally infeasible to directly combine with the community structure in a straightforward Bayesian approach. To tackle this problem, we propose a two-stage strategy that uses a Dynamic Latent Space Model \citep{hoff2002latent,sewell2014dynamic,sewell2016community} to do the clustering at the first stage. That working model should be corresponded to the particular ERG Model you choose as specifically as possible. For example, a distance measure or some covariates should be included if the (hypothesized) true model contains transitivity or homophily effect, respectively. A Latent Space Model that accounts for all dependencies among relational ties plus an appropriate model of community evolution could be expected to recover the clustering change trajectory asymptotically. Then at the second stage, we take the clustering estimates at each time point as given, and further fit a temporal ERG model for actors remaining in each community. In this way, we harness the describing power of more complex models (THERGM versus dynamic SBM) in a timely manner by having the sizes after decomposition under control. For researchers who only care about the inference of (temporal) ERGM, our approach could provide correctness, rather than feasibility, as it could be easily confounded by the hierarchical structure if there are unknown clusters.\\ 
The rest of paper is organized as following: In Section \ref{sec_related}, we briefly state the form of Exponential Random Graph Models. In Section \ref{sec_thergm} we propose our new Hierarchical ERGM for temporal clustered network data. In Section \ref{sec_inference}, we describe the two-phase strategy for inference. In Section \ref{sec_Exp}, various simulation settings for our generative models are used to evaluate the performance of the proposed two-phase strategy. We summarize contributions and discuss limitations as well as future work in Section \ref{sec_Disc}. %In Section \ref{sec_Data}, we apply the proposed procedure to demonstrate the improve of model fit and prediction.

\section{Related Works} \label{sec_related}
\subsection{Exponential Random Graph Models}
Exponential Random Graph Models is a family of statistical models taking the following form of the probability functions
\begin{equation} \label{ergm distribution}
\P_{\theta}\{Y = y\} = \text{exp}\left({\theta}'S(y) - \psi(\theta)\right)
\end{equation}
with arbitrary statistics $S(y)$ for researchers to specify structures of scientific interests. The interpretation of parameters $\theta$ is typically based on the log odds ratio of forming a tie, conditional on the rest of the graph since:
\begin{equation} \label{conditional}
\text{logit}\left( P_{\theta}\{ Y_{i,j}=1 | Y^c_{i,j} \} \right) = {\theta}^{'} c_{i,j}
\end{equation}
where $Y^c_{i,j}=\{Y_{u,v} | \text{ for all } u<v, (u,v)\neq(i,j)\}$ represents all other ties except $Y_{i,j}$, $c_{i,j} = S\left(y^{(ij1)}\right) - S\left(y^{(ij0)}\right)$ is the \textit{change statistic} with $y^{(ij0)}$ and $y^{(ij1)}$ denoting the adjacency matrices with the $(i,j)$th element equal to $1$ and $0$ while all others are the same as $y$.

\subsection{Hierarchical ERGM}
\cite{schweinberger2015local} proposed a hierarchical extension of ERGM by introducing the \textit{local dependence} that breaks down the dependence of random graph $Y$ into subgraphs. Assume there is a partition of the \textit{vertices} V into $K \geq 2$ non-empty finite subsets $V_1, \dots, V_K$, such that the within- and between- neighbourhood subgraphs $Y_{k,l}$ given membership $M$ satisfy
\begin{equation}
\P(Y=y | M=m) = \prod_{k=1}^{K}\P(Y_{k}=y_{k} | M=m)\prod_{l=1}^{k-1}\P(Y_{kl}=y_{kl} | M=m)
\end{equation}

\noindent The within-neighborhood probability measures $\mathbb{P}_{k,k}$ take specific ERGM forms as
\begin{equation}
\P_{\theta_k}(Y_{k}=y_{k} | M=m) = exp\{\theta_k'S_k(y_{k}) - \psi_k(\theta_k) \}
\end{equation}
whereas the between-neighborhood probability measures $\mathbb{P}_{k,l}$ induce independence between subgraphs, and the between-neighborhood ties are assumed to be independent
\begin{equation} \label{eqn:between}
\P(Y_{kl}=y_{kl} | M=m) = \prod_{i \in A_k, j \in A_l} \P(Y_{ij} = y_{ij} | M_i=m_i, M_j=m_j)
\end{equation}

\subsection{Temporal ERGM}
\cite{hanneke2010temporal} proposed a temporal extension of ERGM by making a Markov assumption on the network from one time step to the next.  Specifically,  $Y^t$ is independent
of $Y^1,\ldots,Y^{t-2}$ given $Y^{t-1}$. Taking the first observation $Y^0$ as given, the joint distribution can be factorized:
\begin{equation}
\P(Y^1,Y^2,\ldots,Y^t|Y^0) = \P(Y^t|Y^{t-1})\P(Y^{t-1}|Y^{t-2})\cdots\P(Y^1|Y^0).
\end{equation}
% specify the conditional as an ERG pmf
Given the Markov assumption, ERGM is generalized for evolving networks assuming $Y^t | Y^{t-1}$ admits an ERGM representation. The conditional PDF takes the following form:
\begin{equation}
\P(Y^t | Y^{t-1}, {\boldsymbol \theta}) = \exp\left\{{\boldsymbol \theta}^\prime S(Y^t,Y^{t-1}) - \psi({\boldsymbol \theta},Y^{t-1})\right\}
\label{eqn:tergm}
\end{equation}
The normalizing constant $\psi$ now also depends on $Y^{t-1}$, it is still intractable and the same MCMC approximated MLE techniques apply.\\
\cite{krivitsky2014jrssb} further extended the discrete Temporal ERGM to a separate parameterization of incidence from duration, motived by the reality that social processes and factors of ties forming are very likely not the same as those of ties dissolving.

\subsection{Other Temporal Models}
Another popular class of models for dynamic networks is Stochastic Actor-Oriented Models (SAOM) \citep{snijders2001statistical,snijders2010introduction}. \cite{leifeld2015theoretical} explains how SAOM has an ERGM as its limiting distribution under a specific but rarely applied rate function. TERGM outperforms SAOM substentially when the DSP of SAOM was not met so TERGM is more robust in the sense of model mis-specification. \cite{kim2013nonparametric} models the birth and death of individual groups via a distance dependent Indian Buffet Process and capture the evolution of node group memberships via a Factorial Hidden Markov model.

\section{Temporal Hierarchical ERGM} \label{sec_thergm}
We begin with notations. Consider a temporal series of networks $\{{G}^{(1)}, \ldots, {G}^{(T)}\}$, where ${G}^{(t)} \equiv\{V^{(t)}, E^{(t)}\}$ represents
the network observed at time $t$, consisting a set of \textit{vertices} $V^{(t)}=\{1, \dots, n^{(t)}\}$ and \textit{edges} $E^{(t)}=\{(i,j) | i,j \in V^{(t)} \}$. We assume a fixed number of $K$ communities(or clusters, neighborhoods, blocks), and each \textit{vertex} $i$ has a membership(or color) $m_i^{(t)}=k$ for $k \in {1, \dots, K}$, which may vary at a different time $t'$. The distribution of communities is also allowed to change, as it reflects the overall trend of the system, for example, the rising of ``big data'' in Statistics. In this paper, $G^{(t)}$ is restricted to a binary adjacency matrix $Y^{(t)} = \{Y_{i,j}^{(t)}\}_{1\le i\ne j\le n}$ where
\[
Y_{i,j}^{(t)}=\begin{cases}
1 & \text{if there is an edge between vertices } i \text{ and } j\\
0 & \text{otherwise}
\end{cases}
\]\\
Most of the literature assume that $V=V^{(t)}$ for all $t$ is a fixed set over time, while any dynamically changing network in real world would have nodes entering or leaving the system. It is because of that network structural analysis are concerning about the effects of existing ties on the formation (or dissolution) of future relations, which are implicitly among the same set of nodes. If the nodes change too volatile, it is the reason(s) of entering (leaving) should be modeled. For example, during the first 3 months from the launch of \textit{facebook.com}, what really matters is how to attract people to open accounts, rather than introducing friends' friends. Here we further assume this situation applies to the community level. That is to say, for subsets $V^{(t-1)}_k$ and $V^{(t)}_k$ of \textit{vertices} belonging to community $k$ at any two consecutive time points, the remaining part $V^{(t-1)}_k \cap V^{(t)}_k$ is at least a significant portion of the union $V^{(t-1)}_k \cup V^{(t)}_k$. Together with a finite and pre-fixed number of clusters $K$, we are targeting a gradually evolving system. For streaming graph data, see for example \cite{aggarwal2010stream}.

\subsection{Our model}
Now we propose a class of Temporal Hierarchical Exponential Random Graph Models (THERGM), by assuming 1. $Y^t$ can be decomposed by the current clustering $M^t$ at any time $t$; 2. Markov property of the temporal dependency on $Y^{t-1}$ and also implicitly $M^{t-1}$:
\begin{align*}
\P(\{Y\}^t_1,\{M\}^t_1|Y^{(0)},M^{(0)}) = &\P(Y^{(t)}|Y^{(t-1)},M^{(t)},M^{(t-1)})\cdots \\ 
& \P(Y^{(1)}|Y^{(0)},M^{(1)},M^{(0)}) \P(\{M\}^t_1|Y^{(0)})
\end{align*}
Denote $V^{t}_{k,remain} \equiv V^{t-1}_{k} \cap V^{t}_{k}$ as the set of remaining nodes in group $k$ during the time $t-1$ and $t$, so that ${G}^{t-1,t}_k \equiv\{V^{t}_{k,remain}, E^{t-1,t}_k\}$ and ${G}^{t,t-1}_k \equiv\{V^{t}_{k,remain}, E^{t,t-1}_k\}$ are the subgraphs spanned by the common nodes at time $t-1$ and $t$, respectively. $Y^{t,t-1}_{k}$ and $Y^{t-1,t}_{k}$ are the corresponding adjacency matrices:

\begin{align} \label{eqn_thergm}
\P(Y^{(t)}|Y^{(t-1)},M^{(t)},M^{(t-1)}) = &\prod_{k=1}^{K} \P(Y^{(t,t-1)}_{k} | Y^{(t-1,t)}_{k},M^{(t)},M^{(t-1)}) \nonumber\\
& \P(Y^{(t,t-1)}_{k,new} | M^{(t)},M^{(t-1)}) \nonumber\\
& \prod_{l=1}^{k-1} \P(Y^{(t)}_{kl} | M^{(t)})
\end{align}

\noindent Note the first item in the RHS of equation \ref{eqn_thergm} accounts for the ties associated with remaining nodes in group $k$, which takes the temporal ERGM form:
\begin{align*}
\P(Y^{(t,t-1)}_{k} | Y^{(t-1,t)}_{k},\theta_k) = \exp\left\{{\boldsymbol \theta_k}^\prime S(Y^{(t,t-1)}_{k},Y^{(t-1,t)}_{k}) - \psi({\boldsymbol \theta_k},Y^{(t-1,t)})\right\}
\end{align*}
Note that it is implicitly depending on $M^{(t)}$ and $M^{(t-1)}$.
The second item accounts for the ties associated with nodes newly joined group $k$ from time $t-1$ to time $t$, which can be assumed to follow a \textit{preferential attachment} process \citep{barabasi1999emergence}.
%the probability that an actor has $k$ links has a power-law tail, following 
%\begin{equation}
%P(k) \sim k^{-\gamma}
%\end{equation}
%where $\gamma$ is the scaling factor.
The third item accounts for the between-neighborhood ties, which are assumed to be independent as in equation \ref{eqn:between}. \\

\noindent To complete the THERGM specification, a sensible layer of community evolution is needed. Similar to longitudinal network observations, Markovian assumptions are made to simplify the joint distribution
\begin{align*}
\P(\{M\}^t_0) = \P(M^{(t)}|M^{(t-1)})\cdots\P(M^{(1)}|M^{(0)})\P(M^{(0)}).
\end{align*}
and there are at least three ways to do it:
\begin{itemize}
	%\item \cite{xu2014dynamic,han2014consistent} assume that the cluster memberships $M = M^t$ for $t=0,1,\ldots,T$ do not change over the observation periods, but only the \textit{class connection probability matrix} of their assumed underlying stochastic block model, $P_{kl}^t$, may be different. Our model can adopt the same linear Gaussian process approach of modeling $\{P_{kl}\}^T_1$ for the between-neighborhood probability.
	\item \cite{yang2011detecting,sewell2016community} use a stationary transition matrix $\{\beta_{hk}\}$, to denote the probability of a node in cluster $h$ at time $t-1$ to become a member of cluster $k$ at next step $t$.
	\item \cite{xing2010aoas,ho2011evolving} impose a logistic normal prior with a State Space model to model the trajectory.
	\item \cite{ghasemian2015detectability} propose that with probability $\eta$, each node keeps its label from one time step to the next, otherwise it chooses a new label $m_i^t$ from the prior $q_r$. The transition probability for community memberships is
	\begin{equation}
	P( M^t \,|\, M^{t-1} )= \prod_{i} \left( \eta\, \delta_{m_i^t,m_i^{t-1}} + (1-\eta) q_{m_i^t}\right) \enspace,
	\label{eq:tran_prob}
	\end{equation}
\end{itemize}

We adopt the first way of modeling, and implicitly assume that the community evolution does not depend on the realized networks. See \textit{latent feature propagation} model of \cite{heaukulani2013dynamic}, to capture the phenomenon that observed social interactions from the past can also affect future unobserved structure. The proposed dynamic process can be illustrated in Figure \ref{fig_mechanism}:
\begin{figure}
	\includegraphics[width=0.5\textwidth,height=0.25\textheight]{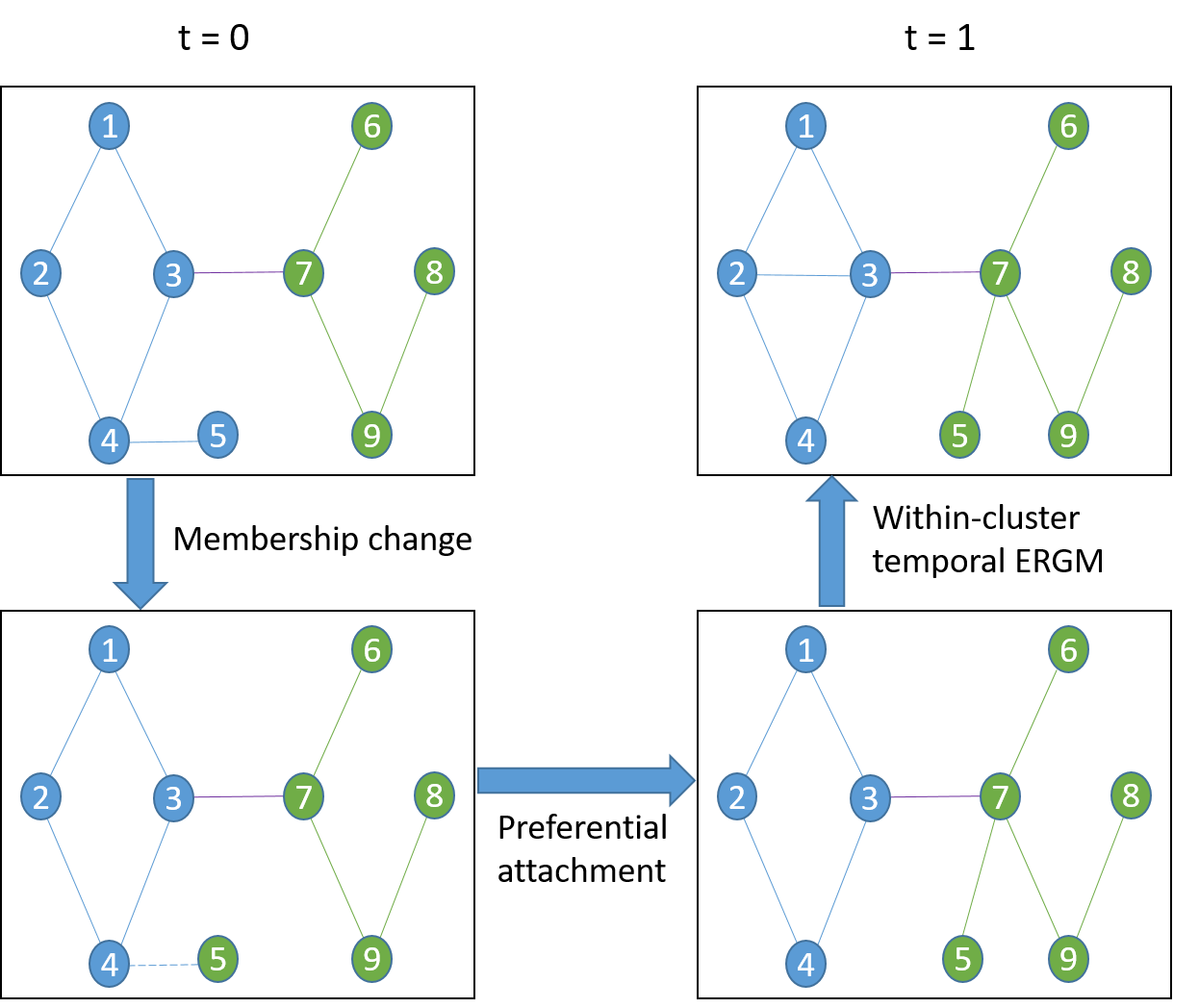}
	\caption{Community evolution as well as (within-cluster) link forming illustrated in one step. Node $5$ was belong to group one (blue) at $t0$ but changed to group two (green) at $t1$, reflecting an overall trend of the rising group two. It chose which node to connect in the new group with the probability proportional to that node's degree (i.e. Preferential attachment). A link between $5$ and $7$ is formed as node $7$ has the most existing edges, while the link between $5$ and $4$ dissolved as $5$ left group one. Finally, the remaining nodes in each group form links according to their specific ERGM formula. In this case, a link between $2$ and $3$ is formed as it completes two triangles (transitivity).}
	\label{fig_mechanism}
\end{figure}

\section{Model inference} \label{sec_inference}
A straightforward full Bayesian approach to estimate THERGM parameters is not practical. At the clustering level, the sample space of membership $M$ is $K^n$ where $K$ is the number of blocks and $n$ is the number of nodes. At the within-block TERGM part, the sample size of edge variable $Y_k$ is $2^{n_k \choose 2}$ where $n_k$ is the number of nodes in $k$th block (each pair of nodes can have a link present or absent in an undirected binary network). So both parts need MCMC or other sampling methods to do the approximation, directly combining them makes the problem intractable. In this section, we propose a two-stage strategy to tackle this problem.

\subsection{Clustering with a working model} \label{twostage_inf}
Now, we temporarily jump out of the ERGM paradigm and take a look at another class of statistical models that mainly embracing the classical latent factor idea. Instead of explicitly modeling dependence, the Latent Structure Models (LSM) postulate latent nodal variables $Z$ and conditional independence of $Y_{i,j}$ given those variables $Z_i$ and $Z_j$.\\
\cite{hoff2002latent} introduced the concept of unobserved "social space" within which each node has a position so that a tie is independent of all others given the unobserved positions of the pair it connects to:
\begin{equation}
P(Y=y | Z, X, \beta) = \prod_{i \neq j}P(Y_{i,j}=y_{i,j} | x_{i,j}, z_i, z_j, \beta)
\end{equation}
where $X$ are observed covariates, and $\beta$ and $Z$ are parameters and positions to be estimated. \cite{handcock2007cluster} took a subclass Distance Models where the probability of a tie is modeled as a function of some measure of distance between the latent space positions of two nodes:
\begin{equation}
\text{logit}\{P(Y_{i,j} = 1 | x_{i,j}, z_i, z_j, \beta)\} = \beta_0^{T}x_{i,j} - \beta_1|z_i - z_j|
\end{equation}
with restriction of $\sqrt{\frac{1}{n}\sum_{i}{|z_i|}^2} = 1$ for the identification purpose. Then they imposed a finite mixture of multivariate Gaussian distribution for $z_i$ to represent clustering:
\begin{equation}
z_i \overset{\text{iid}}{\sim} \sum\limits_{k=1}^{K} \lambda_k \textbf{MVN}_{d}(\mu_k, \sigma_k^{2}\textbf{I}_d)
\end{equation}
where non-negative $\lambda_k$ is the probability that an individual belongs to the $k$th group, with $\sum\limits_{k=1}^{K}\lambda_k = 1$.\\
\cite{sewell2016community} proposed a dynamic extension of LSM, assuming at each time point, each node belongs to one of a fixed number $K$ of clusters, and the membership may change over time. Denote the latent position of node $i$ at time $t$ as $Z_{i}^{t}$ and the corresponding membership as $M_{i}^{t}$. Let $Z^t = ({Z_{1}^t}', \ldots, {Z_{n}^t}')$ and $M_t = (M_{1}^t, \ldots, M_{n}^t)$, with the additional assumption that given the latent positions $Z^t$, $Y^t$ and $Y^s$, $s \ne t$, are conditionally independent, the decomposition follows:
\begin{equation}
P(\{Y^t, Z_t, M_t\}_1^T) = \prod_{t=1}^{T} \P(Y^t | Z^t) \times \P(\{Z_t, M_t\}_1^T)
\end{equation}

Two assumptions are made on the latent positions and the cluster assignments. First, the cluster assignments are assumed to follow a Markov process, i.e.,
\begin{equation}
M_i^t | M_i^1,\ldots,M_i^{t-1} = M_i^t | M_i^{t-1}
\end{equation}
Second, given the current cluster assignment and all previous cluster assignments and latent positions, we assume the current latent positions depend only on the previous latent position and the current cluster assignments, i.e.,
\begin{equation}
Z_i^t | Z_i^1,\ldots,Z_i^{t-1}, M_i^1,\ldots,M_i^{t} = Z_i^t | Z_i^{t-1},M_i^t
\end{equation}

\subsection{Identification and label switching}
The cluster label is identified up to a permutation. We adopted the hamming error definition to define a correct membership, which is the minimal clustering errors over a permutation $\pi$ of the set $\{1, 2, \ldots, K\}$. See section $2.6$ of \cite{jin2015score} for details. Here we want to make explicitly an assumption that the label switching problem between two consecutive time steps can be solved by the permutation of the hamming error. This assumption could be easily satisfied when the temporal changes are not very dramatic.

\section{Experiments} \label{sec_Exp}
\subsection{Simulation setting}
We use Dynamic Stochastic Block Model (DSBM) \citep{matias2017dynamic} as a comparison.
The dissolving rate is set to $10$ time steps. The initial density is set to be $0.1$ and kept roughly stable if without the transitivity effect. Other parameters are ... All the simulations are carried out using the R package \textit{Simulator} \cite{bien2016simulator}.\\
One example of the simulated longitudinal networks, with some summary statistics are shown in Figure \ref{fig_sample}:
\begin{figure*}
	\includegraphics[width=\textwidth,height=0.3\textheight]{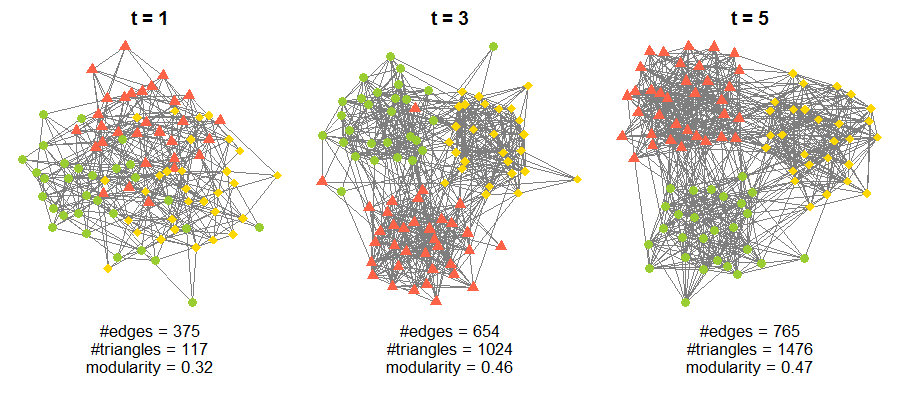}
	\caption{For simplicity, we only plot networks at time $t=1$, $t=3$ and $t=5$, while the whole dynamic network data contains 5 networks.}
	\label{fig_sample}
\end{figure*}

\subsection{Transition matrix estimate}
\begin{figure}
	\includegraphics[width=0.5\textwidth]{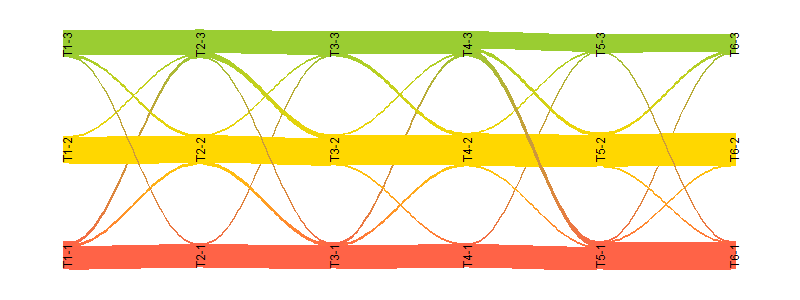}
	\caption{This river plot illustrates a setting of quickly and imbalanced changing transition matrix in our simulation. Each bar (river) denotes one community and each curve (flow) denotes there are nodes shifting membership. A label `T\textit{x}-\textit{y}' represents time point \textit{x} for group \textit{y}. The green (yellow) bar is getting thinner (thicker) reflects there are more (less) nodes flow into group three (two).}
	\label{fig_river}
\end{figure}

%use the texreg package to draw
In a lot of applications, the overall trend of the communities change, for example, the rising of a certain group. The following \textit{river plot} is an illustration of how nodes ``flow'' from one ``river'' into another. The height of each bar, in other words, the width of each river, represents the relative size of that community, illustrated in Figure \ref{fig_river}.

\subsection{Evolving membership estimate}
\begin{figure*}
	\includegraphics[width=\textwidth,height=0.5\textheight]{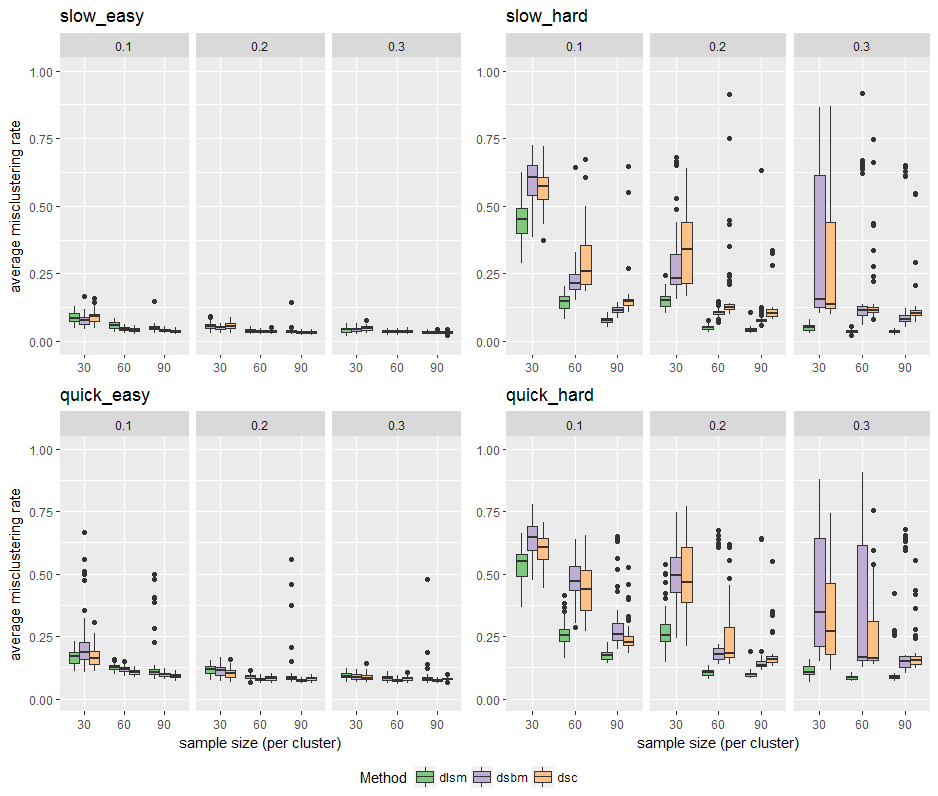}
	\caption{Average mis-clustering rates as sample size and transitivity strength change, in four scenarios respectively.}
	\label{fig_avgmis}
\end{figure*}

We investigate the effects of sample size (per cluster) and the transitivity strength under four scenarios, which are the combination of two factors: the time transition rate (slow v.s. quick) and within cluster density (easy when the difference of within- and between- cluster density is large). The results in Figure \ref{fig_avgmis} indicate that when it is slow and easy, all methods perform well, while under quicker or harder situations, the dynamic latent space model perform better as it is a better working model.

\begin{figure*}
	\includegraphics[width=\textwidth,height=0.4\textheight]{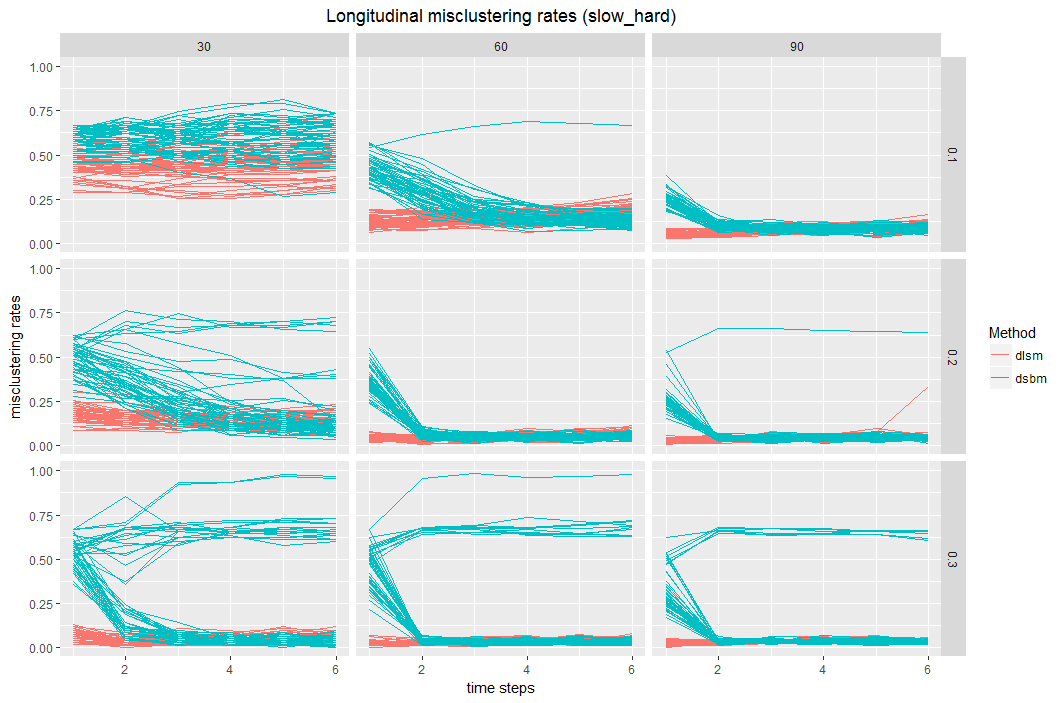}
	\caption{spaghetti plot type of mis-clustering rate along time under different sample sizes.}
	\label{fig_longmis}
\end{figure*}

If we dig into the longitudinal mis-clustering rates, we can see that DSBM has a non-negligible probability that error rate could not be controlled, while DLSM can as shown in Figure \ref{fig_longmis}.

\subsection{Cluster specific ERGM estimate}
\begin{figure*}
	\includegraphics[width=\textwidth]{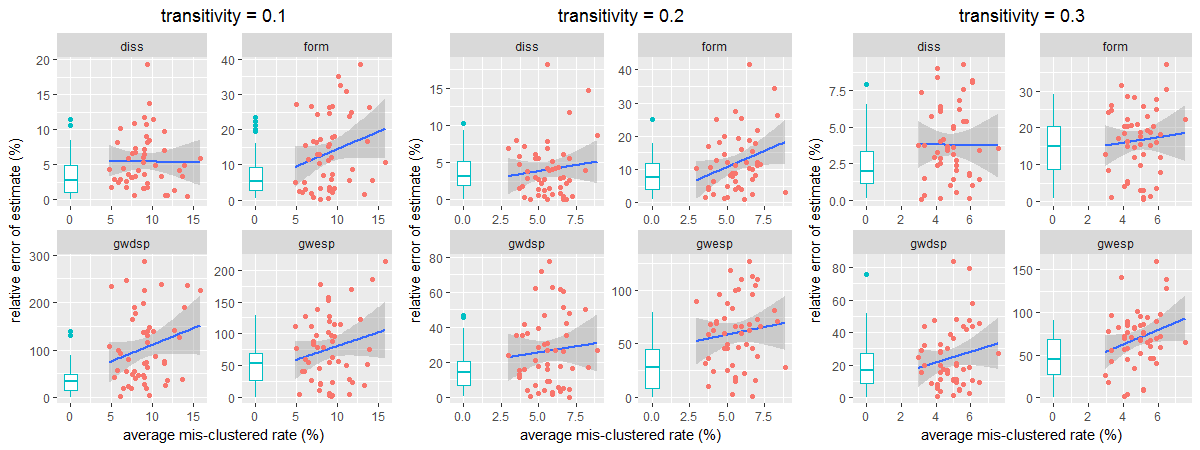}
	\caption{The green boxplot centered at $0$ is the TERGM estimates of true clustering (no estimation error). The red dots are relative error of estimates, conditioning on the estimated membership.}
	\label{fig_parest}
\end{figure*}

The TERGM estimates conditioned on the membership estimates in the first step is carried out by \cite{krivitsky2014jrssb}. Results illustrated in Figure \ref{fig_parest}. %Bootstrapped Pseudolikelihood \cite{btergm_package}. It is much faster to get a 

\subsection{Goodness of fit}
\begin{figure}
	%\centering
	\begin{minipage}[b]{0.4\textwidth}
		\includegraphics[width=\textwidth]{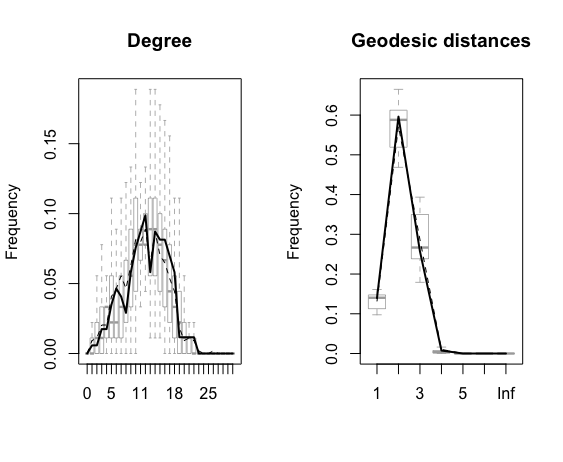}
	\end{minipage}
	\begin{minipage}[b]{0.4\textwidth}
		\includegraphics[width=\textwidth]{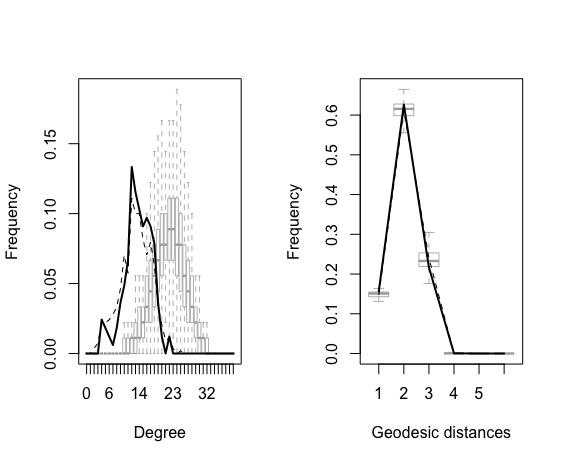}
	\end{minipage}
	\caption{While they both fit Geodesic distances well, the estimated THERGM has a much better fit to the Degree distribution.}
	\label{fig_gof}
\end{figure}

To answer the natural question that if my working model is already good enough, why bother fitting within-cluster TERGMs, we show Figure \ref{fig_gof}.

\subsection{Link prediction}
\begin{figure}
	\includegraphics[width=0.5\textwidth]{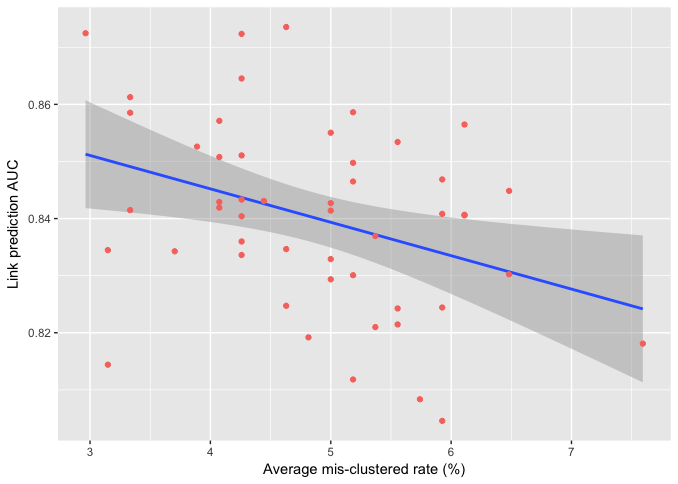}
	\caption{Link prediction AUC decreases as the mis-clustering rate increases.}
	\label{fig_auc}
\end{figure}

As in a lot of machine learning applications, the out-of-sample prediction accuracy is the most trustworthy metric. So we conducted a one time step out-of-sample prediction and calculated the Area Under the ROC curves (AUC) in Figure \ref{fig_auc}.

\section{Discussion and future work} \label{sec_Disc}
As the experiments show, the clustering performance is not perfect, and the mis-clustered nodes have huge impact on the second stage TERGM inference. We are currently working on an augmented Bayesian method to jointly infer the clustering and local structures. The tool we use is the newly developed probabilistic programming language \textit{Edward} \cite{tran2016edward,tran2017deep}, which integrates the \textit{Google Tensorflow} \citep{abadi2016tensorflow} framework with its capability to utilize GPU.

\bibliographystyle{abbrv}
\bibliography{}
\end{document}